\documentstyle[12pt]{article}
\begin{document}

\title{On the proximity relation between two surface-melted clusters involved in
inter-cluster mass-transfer }
\date{October 27, 2000}
\author{F. Despa and R.S. Berry \\
%EndAName
Department of Chemistry,\\
The University of Chicago,\\
Chicago, Illinois 60637}
\maketitle

\begin{abstract}
We explore the way free particles produced by dissociating ``particle-hole
pairs'' on a surface-melted cluster can be transferred to a second, nearby
surface-melted cluster. This mass transport is based on an inter-cluster
direct transfer mechanism of the particles. We found that in this particular
case one cluster may grow at the expense of another, obeying a temporal
power law with the exponent $1/2$ for the average radius $(R\sim t^{1/2})$.
The change from the expected universal power law $(R\sim t^{1/3})$ is a
consequence of the proximity relation between these two clusters which lead
to enhance the effective transport rates.
\end{abstract}

It is widely agreed that, under certain conditions, clusters may exhibit
more than two phaselike forms coexisting in dynamic equilibrium.\cite
{nauchitel,cheng1,cheng2,kunz1,kunz2} For example, $Ar_{55}$ is expected to
show, depending on the temperature range, coexistence of solidlike,
homogeneously melted forms and surface-melted either in an ensemble at any
instant or in the time history of a single cluster.\cite{kunz1,kunz2} It is
also equally possible, e.g., for the solid form to be in equilibrium with
only the surface-melted form and for the surface-melted form to coexist at
higher temperatures with the liquid. There may or may not be a temperature
range in which the surface-melted form is the only stable species.

The surface-melting phenomenon is usually exhibited in bulk matter and in
large clusters (with about $50$ or more particles, usually) at temperatures
or energies a little below the temperatures or energies of homogeneous
melting. This process involves local minima on the potential energy surface
of the system corresponding to one or a few of the particles from the
surface layer moving into local sites above the surface. The promoted
particles are rather free and float on the cluster surface while all the
other particles in the surface layer undergo large-amplitude, anharmonic
oscillations. The numerical experiments indicate that the surface is
liquidlike and all the particles of the surface, including the floaters, do
permute among themselves without involving inner layers of the cluster.

The statistical mechanical underpinning of these findings allow for one
other kind of behavior yet to be observed. This would be a dynamical process
within which free particles produced by dissociating ``particle-hole pairs''
on a surface-melted cluster are allowed to recombine with available
vacancies on a second, nearby surface-melted cluster. The inter-cluster
direct mass transfer arisen in this way involves pairs of surface-melted
clusters in a dense cluster ensemble. The occurrence of this process is
limited by the energetic need to drive the particle flow and by an obvious
requirement on the nearest-neighbor distance. We assume the surface-melted
clusters are close to each other such that the direct transfer of the
particles makes sense. The escape energy for the itinerant particle is
achieved by thermal excitation and possible interaction with the surface
vacancies on the neighbor cluster. The probability of transfer is high if
the solid angle in the direction of jump is large. Therefore, the net flow
of particles proceeds from the small cluster, with large curvature, to the
large cluster in the pair, which has smaller curvature. Note that, an
increase of the nearest-neighbor distance between surface-melted clusters
over a critical value may lead to change the transport mechanism and sets in
crossover phenomena.

The investigation of the direct mass transfer between neighboring
surface-melted clusters is of both theoretical and experimental interest.
First of all, this is a useful way to increase our understanding of phase
separation, coarsening\cite{koch} and crossover phenomena.\cite
{lebowitz,cumming,tokuyama,coulon} The process can also have some relevance
in describing kinetic transformations of the cluster-assembled materials.
Second, the direct transfer mechanism of particles between surface-melted
clusters offers the prototype of grain growth which deviates from the
expected universal power law $(S\sim t^{2/3})$ of the classical Ostwald
ripening theory.\cite{koch} Generalization for solid clusters is
straightforward.

Making reference only to the cluster-pair problem appears as a strong
limitation in the broad context as declared above. Usually, more than two
individuals in the entire cluster ensemble may act in this process. We
believe that, for a dense cluster distribution, the chief particle transfer
occurs, anyway, between nearest neighboring clusters, along the shortest
separation distance where the particle concentration gradient is greatest
possible.

It is the purpose of this work to explore the dynamics of such a process.
The modelling starts by assuming that the separation distance between the
surface-melted clusters involved in this process, say $\xi $, is of order of
magnitude of the characteristic diffusion length $l$ $\left( \xi \simeq
l\right) $. In this context, the particles released from one cluster may
reach the second, nearby cluster by performing an inter-cluster direct jump.
Next, we assume that the surface-melting process produces an amount of
``particle-hole pairs''at the surface of each cluster. The number $M$ of
``particle-hole pairs'' is in direct proportion with the number of atoms on
the cluster surface of area $S$ and depends on the temperature $T$ 
\begin{equation}
M\simeq S\sigma \exp \left( -\frac{E_{p-h}}{k_{B}T}\right) \;\;\;,
\label{one}
\end{equation}
where $\sigma $ stands for the surface particle density, $E_{p-h}$ is the
energy to excite a particle above the melted surface (the ''particle-hole
pair'' creation energy) and $k_{B}$ has the usual meaning. The floating
particles perform continuous permutations (atomic scale jumps) among
energetically equivalent sites (vacancies) on the cluster surface with the
frequency 
\begin{equation}
\upsilon _{0}=\upsilon _{p-h}\exp \left( -\frac{E_{0}}{k_{B}T}\right) \;\;\;,
\label{two}
\end{equation}
where $\upsilon _{p-h}$ is the vibration frequency of the ''particle-hole
pair'' and $E_{0}$ is the energy cost of the ''on-site'' jump. The
equivalent sites on the surface of this cluster are labelled by $i$, $i=%
\overline{1,M}$.

According to numerical experiments,\cite{kunz1,kunz2} the floating particles
are rather free and get easily dissociate from the cluster surface. We
suppose these particles can jump to the nearest neighbor cluster, over the
separation distance $\xi $, where may recombine with vacancies of the host
melted surface. The equivalent sites on the surface of the second, nearby,
cluster (of surface area $S^{*}$) are labelled by $k$, $k=\overline{1,M^{*}}$
with $M^{*}$ given by 
\begin{equation}
M^{*}\simeq S^{*}\sigma \exp \left( -\frac{E_{p-h}}{k_{B}T}\right) \;\;\;.
\label{three}
\end{equation}
The ''inter-site'' transfer frequency is denoted by
\begin{equation}
\upsilon \simeq \upsilon _{p-h}\exp \left( -\frac{E_{d}}{k_{B}T}\right)
\;\;\;,  \label{four}
\end{equation}
where $E_{d}$ stands for the energy cost to dissociate the ''particle-hole
pair''.

We place now one cluster at $x$ and the other at $x+\xi $ and apply to this
mass-transfer process a kinetic approach based on a system of master
equations which describe both permutations and escapes of itinerant
particles.\cite{despa} Accordingly, the rate of change of the particle
number at every site ''$i$'' on the melted surface of the cluster placed at $%
x$ may be written as 
\begin{eqnarray}
\frac{\partial }{\partial t}n_{i}(x,t) &=&\upsilon _{0}\sum_{i\neq
j=1}^{M}n_{j}(x,t)-\upsilon _{0}\sum_{i\neq j=1}^{M}n_{i}(x,t)+  \label{five}
\\
&&\upsilon \sum_{k=1}^{M^{*\prime }}n_{k}(x+\xi ,t)-\upsilon
\sum_{k=1}^{M^{*\prime }}n_{i}(x,t)\;\;\;\;.  \nonumber
\end{eqnarray}
After some simplifications, this becomes 
\begin{eqnarray}
\frac{\partial }{\partial t}n_{i}(x,t) &=&\upsilon _{0}\sum_{i\neq
j=1}^{M}\left[ n_{j}(x,t)-n_{i}(x,t)\right] +  \label{six} \\
&&\upsilon \sum_{k=1}^{M^{*\prime }}n_{k}(x+\xi ,t)-\upsilon M^{*\prime
}n_{i}(x,t)\;\;.  \nonumber
\end{eqnarray}
The first term on the right hand side of the equation accounts for the
movement of the particles on the surface of the cluster (the number of
floaters is conserved by this movement) while, the second term includes all
transfer possibilities of the particles to the second, nearby cluster
(naturally, this transfer changes the particle number at each site $i$).
Obviously, those sites on cluster surfaces which do not face each other are
excepted from the calculus. Therefore, $M^{*\prime }$ counts the effective
available sites on the surface of the cluster placed at $x+\xi $ which can
be reached by a direct jump, $M^{*\prime }=fM^{*}$ where $f$ is the weight
of the geometrical obstruction (roughly, $f\simeq \frac{1}{2}$). Itinerant
particles, subject of ''inter-site'' jumps, distribute themselves, quickly,
over the whole surface area by ''on-site'' jumps (with the frequency $%
\upsilon _{0}$, $\upsilon _{0}\gg \upsilon $). In this way, all the
''particle-hole pairs'' get involved in the transport process. By summing up
over all sites ''$i$'' in $\left( 6\right) $ we obtain 
\begin{equation}
\frac{\partial n}{\partial t}=\upsilon _{b}n^{*}-\upsilon _{f}n\;\;\;,
\label{seven}
\end{equation}
which is the transport equation for the particle number $n(x,t)=%
\sum_{i=1}^{M^{\prime }}n_{i}(x,t)$ with $n^{*}(x+\xi
,t)=\sum_{k=1}^{M^{*\prime }}n_{k}(x+\xi ,t)$. $\upsilon _{f}$ and $\upsilon
_{b}$ stand for the forward and backward effective transition rates
\begin{eqnarray}
\upsilon _{f} &\simeq &\frac{1}{2}M^{*}\upsilon =\frac{S^{*}\sigma }{2}%
\upsilon _{p-h}e^{-\frac{E}{k_{B}T}}  \label{eight} \\
\;\upsilon _{b} &\simeq &\frac{1}{2}M\upsilon \;=\;\frac{S\sigma }{2}%
\upsilon _{p-h}e^{-\frac{E}{k_{B}T}}\;\;,  \nonumber
\end{eqnarray}
where $E=E_{d}+E_{p-h}$. The same set of equations $\left( 5-7\right) $
apply for the cluster placed at $x+\xi $ and one get straightforwardly the
transport equation for the particle number $n^{*}(x+\xi
,t)=\sum_{k=1}^{M^{*\prime }}n_{k}(x+\xi ,t)$ as 
\begin{equation}
\frac{\partial n^{*}}{\partial t}=\upsilon _{f}n-\upsilon _{b}n^{*}\;\;\;.
\label{nine}
\end{equation}

By looking at equations $\left( 7-9\right) $ we may infer that the net flow
of itinerant particles occurs in the direction of the large cluster where
the solid angle of the particle jump is large. For example, we may consider
that at the initial moment of time $t=0,$ the cluster placed at $x+\xi $ is
larger by comparing with the other, $S^{*}>S$, which means that $\upsilon
_{f}>\upsilon _{b}$. Consequently, the cluster placed at $x+\xi $ starts
growing on the expense of the other which shrinks in time. The corresponding
variation in time of the cluster surfaces $S^{*}$ and $S$ have to be in
direct proportion with the instant numbers of particle concentrations $n$
and $n^{*}$ given by kinetic equations $\left( 7-9\right) $ and should
depend on the rate of spreading over the cluster surface (this is given by
the transfer frequency $\upsilon _{0}$, subject of eq. $2$). The fact can be
expressed by the following set of coupled differential equations 
\begin{eqnarray}
\frac{1}{A}\frac{dS^{*}}{dt} &=&n^{*}\left( t\right) ;\;\frac{1}{A}\frac{dS}{%
dt}=-n\left( t\right)   \label{ten} \\
\frac{1}{B}\frac{\partial n^{*}}{\partial t} &=&nS^{*}-n^{*}S;\;\frac{1}{B}%
\frac{\partial n}{\partial t}=n^{*}S-nS^{*}\;\;\;,  \nonumber
\end{eqnarray}
where the forward and backward transition rates $\upsilon _{f,b}$ in above
were replaced by $\left( 8\right) $ and use has been made of  eq. $\left(
2\right) $ which render for the constants $A$ and $B$ 
\begin{equation}
A=\frac{\upsilon _{p-h}}{\sigma }e^{-\frac{E_{0}}{k_{B}T}};\;\;B\simeq \frac{%
\sigma \upsilon _{p-h}}{2}e^{-\frac{E}{k_{B}T}}\;\;\;.  \label{eleven}
\end{equation}

The evolution in time of cluster sizes depends on the degree of melting of
their surfaces by the corresponding energy cost $E=E_{p-h}+E_{d}$ and by the
characteristic (anharmonic) oscillation frequency $\upsilon _{p-h}$. In Fig.
1, we displayed the evolution in time of the surface area for the growing
cluster, $S^{*}\left( t\right) $, following from the inter-cluster direct
transfer mechanism of particles with the growth law given by eqs. $\left(
10\right) $ (the curve $a$). We may roughly approximate the growth law goes
linearly $\left( S^{*}\sim t\right) $, which means the increase of the
average radius of the cluster obeys a power law with the exponent equal to $%
1/2$ .\cite{marqusee}

We may observe that, actually, the set of eqs. $\left( 10\right) $ are
general equations and can be used to describe the effective mass-transfer in
any situation involving pairs of clusters which are not necessarily
surface-melted but satisfy the appropriate proximity requirement $\left( \xi
\simeq l\right) $. In the general case, $E$ is simply the dissociation
energy for an atom at the cluster surface and $\upsilon _{p-h}$ becomes the
surface vibration frequency $\left( \upsilon \right) $.

The result obtained above refers to the mechanism of direct transfer of
particles between neighboring clusters. In principle, the process can switch
to an asymptotic transport regime by increasing the separation distance
between the two clusters. For widely separated clusters, the latter case
resumes to an ideal one, where there is only one cluster growing directly
from solution. The cluster growth obeys, in this case, the well-known power
law 
\begin{equation}
R\sim t^{1/3}\;\;,  \label{twelve}
\end{equation}
for the average radius and is known as the Ostwald ripening process.\cite
{koch} The growth of surface area goes as $\sim t^{2/3}$, in this case. For
comparison, we displayed in Fig. 1 the growth law obeying the $2/3$ power
law (see the curve $b$).

The asymptotic regime occurs naturally in our model by letting the
separation distance $\xi $ between clusters go to infinity $\left( \xi \gg
l\right) $ which allows the itinerant particles to proceed by a random walk.
By looking above we can see that this crossover between the direct particle
transfer regime and the asymptotic one can be set up by transforming the
kinetic equation $\left( 7\right) $ into a diffusion-like type of equation.
This transformation requires $\upsilon _{f}=\upsilon _{b}\equiv \upsilon $
and a Taylor expansion up to second order of the concentration functions.
The procedure allows to obtain the diffusion coefficient $\left( D\sim
\upsilon l^{2}\right) $ which is the constant associated with the diffusion
through the background matrix at large distances from the shrinking cluster.
The rate of growth is then given by $\left( 12\right) $. Of course, the
transition to the asymptotic regime proceeds gradually, being characterized
by various transient values of the time exponents in the range $1/2$ through 
$1/3$.\ Finally, we can note that the main factor of delaying the setting up
of the asymptotic regime is the proximity relation between the two clusters
involved in the mass-transfer.

As one can see, the limiting regime $\left( R\sim t^{1/2}\right) $
identified here in the cluster growth process, namely the inter-cluster
direct transfer of particles, has no relevant statistical aspects: this is a
particle transfer process between two neighboring surface-melted clusters
(therefore, essentially a cluster-pair problem). The statistical size
distribution of clusters may change only irrelevantly at its two ends, where
the small and the large clusters are placed. The statistical aspects are
significant, in the sense that the distribution is much and fastly affected,
in the transient regimes intervening between the two limiting ones ($R\sim
t^{1/2}$, $R\sim t^{1/3}$) described above.

{\Large Figure caption}

Fig. 1 - The growth of the surface area of a surface-melted cluster by the
inter-cluster direct transfer mechanism (a) in comparison with the
asymptotic regime of the Ostwald ripening process (b). The parameters
employed in the present computation are: $\upsilon _{p-h}=2\;10^{3}\;s^{-1}$%
, $E_{0}=0.6\;eV$, $E=0.82\;eV$, $kT=0.05\;eV$ and $\sigma =1/4\;\AA ^{-2}$.

\end{document}